\documentclass{ws-procs975x65}
\usepackage{hyperref}
\usepackage{url}
\usepackage[lflt]{floatflt}
\newcommand{\Mpc}{{{\rm Mpc}^{-1}}}

\newcommand{\nS}{{{n_{\rm S}}}}

\newcommand{\id}{{{\rm d}}}

\newcommand{\kB}{{{k_{\rm B}}}}

\begin{document}

\title{Mixing of blackbodies: Increasing our view of inflation to 17
  e-folds with spectral distortions from Silk damping}

\author{Rishi Khatri}

\address{Max Planck Institut f\"{u}r Astrophysik, Karl-Schwarzschild-Str. 1
    85741, Garching, Germany\\
khatri@mpa-garching.mpg.de}
\bodymatter
\begin{abstract}
Silk damping in the early Universe, before and during recombination,  erases anisotropies in the cosmic
microwave background (CMB) on small scales. This power, which disappears from
anisotropies, appears in the monopole as $y$-type, $i$-type and $\mu$-type
distortions. The observation of the CMB spectral distortions will thus make
available to us the information about the primordial power spectrum on scales
corresponding to the comoving wavenumbers $8\lesssim k \lesssim 10^4~\Mpc$ increasing our total view of inflation, when combined with CMB
anisotropies, to span $17$ e-folds. These distortions can be understood
simply as mixing of blackbodies of different temperatures and the
subsequent comptonization of the resulting distortions.
\end{abstract}

\section{A view of inflation spanning 17 e-folds}
\begin{floatingfigure}[l]{9cm}
\includegraphics[width=8.5cm]{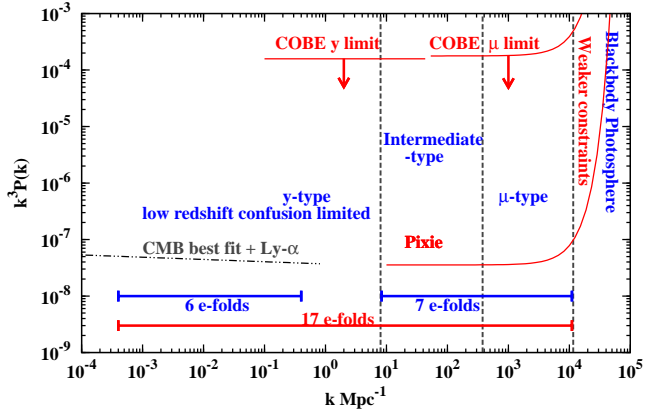}
\caption{\label{fig1} A possible view of $17$ e-folds of inflation.}
\end{floatingfigure}
Successes of CMB experiments such as COBE \cite{cobe,cobedmr}, WMAP
\cite{wmap}, SPT \cite{spt}, ACT \cite{act} and many others\footnote{see
  NASA LAMBDA website for a complete list,
  \url{http://lambda.gsfc.nasa.gov/product/expt/}} has ushered in the era
of precision cosmology. Combination of CMB anisotropies and Lyman-$\alpha$
constraints from SDSS \cite{ly2006,ssm2006} gives information about the
primordial anisotropies on scales from the horizon size today, comoving
wavenumber  $k\sim
4\times 10^{-4}~\Mpc$,  to scales of $k\sim 0.2~\Mpc$, a view of inflation
spanning $\sim 6$ e-folds. 

Any energy released into CMB at redshifts $z\gtrsim 2\times 10^6$ is very
quickly thermalized by the combined action of double Compton scattering and
bremsstrahlung, which create photons at low frequencies, and Compton
scattering which redistributes these photons\cite{sz1970,dd1982} to maintain  Bose-Einstein
spectrum  with occupation number
$n(x)=1/\left(e^{x+\mu}-1\right)$, where  $\mu$ is the chemical potential parameter, 
 dimensionless frequency $x=h\nu/\kB T$,  h is
Planck's constant, $\kB$ is Boltzmann's constant. The parameter $\mu$ is driven to zero because of the creation
of photons by double Compton scattering and bremsstrahlung. At redshifts
$z\lesssim 2\times 10^6$ photon creation becomes very inefficient and any
energy injected is comptonized to $\mu$-type ($2\times 10^5\lesssim
z\lesssim 2\times 10^6$), intermediate ($i$-type)\cite{ks2012b} between $\mu$ and
$y$-type ($1.5\times 10^4\lesssim
z\lesssim 2\times 10^5$)  and $y$-type\cite{zs1969}  ($ 
z\lesssim 1.5\times 10^4$) distortions. 
Photon diffusion damps the perturbations in the primordial plasma on small
scales \cite{silk,Peebles1970,kaiser}, wavenumbers $k\approx 10^4~\Mpc$ getting damped at
$z\approx 2\times 10^6$ and $k\approx 8~\Mpc$ at $z\approx
1.5\times 10^4$. Observations of $\mu$-type and $i$-type distortions will
thus allow us to measure the amplitude and spectral index of the primordial
power spectrum on these very small scales, see 
Fig. \ref{fig1}. $y$-type distortions due to Silk damping are
indistinguishable and are of much smaller amplitude compared to those  
created later during reionization  and are thus not a very useful probe of
primordial power spectrum \cite{cks2012}, hence the gap in Fig. \ref{fig1}
at $k\lesssim 8~\Mpc$. Also shown are the current limits from COBE
\cite{cobe} which constrained at $2-\sigma$ level to $y\lesssim 1.5\times
10^{-5}$ and $\mu\lesssim 9 \times 10^{-5}$. Proposed experiment Pixie
\cite{pixie} will be able to detect the spectral distortions for the WMAP
values of the amplitude of primordial power spectrum\cite{wmap} and
spectral index $\nS=0.96$. Spectral distortion can thus deliver to us $7$ additional
e-folds, reaching $k\approx 10^4~\Mpc$,  giving us a view of inflation
spanning  a total of 17 e-folds 
when combined with CMB anisotropies.

\section{Silk damping as mixing of blackbodies}

\begin{figure}
\includegraphics[width=14cm]{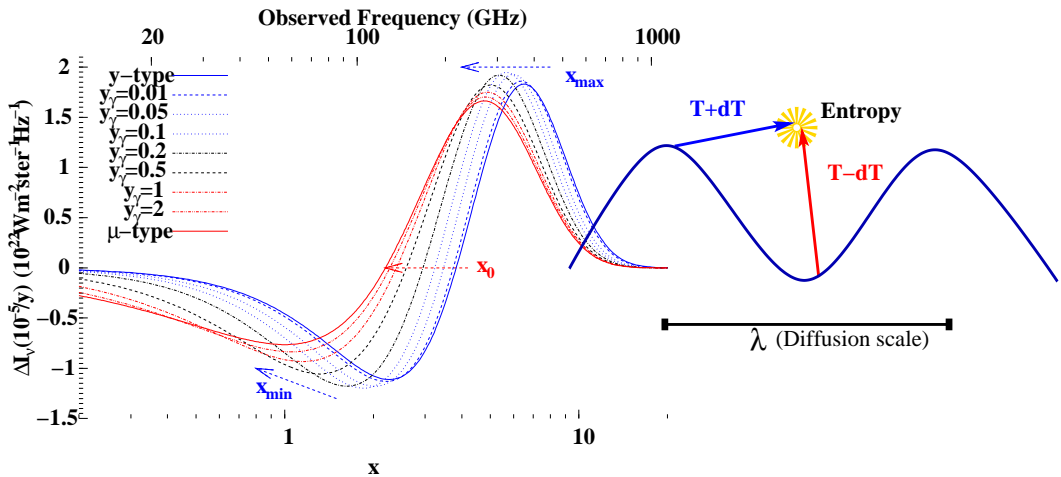}
\caption{\label{fig2}Silk damping can be understood in terms of mixing of
  blackbodies creating $y$-type distortion which comptonizes to $i$-type
  and $\mu$-type distortions. Figures from Refs. ~\protect\refcite{ks2012b} and ~\protect\refcite{ksc2012b}.}
\end{figure}

CMB spectral distortions from Silk damping were previously calculated by
Refs.~\refcite{sz1970b,hss94,daly1991} who, however, overestimated the energy going
into spectral distortions. The calculation of spectral distortions becomes
straightforward once we realize that photon diffusion in the early Universe
mixes photons from parts of the Universe with different temperature, see
right panel in Fig. \ref{fig2}. We
thus expect a $y$-type distortion from the mixing of blackbodies
\cite{zis1972,cs2004,cks2012,ksc2012b} which can subsequently comptonize into $\mu$ and
$i$-type distortions\cite{ks2012b} (at $z\gtrsim 1.5\times 10^4$). This
evolution of $y$-type into $i$-type and 
$\mu$-type distortions is shown in Fig. \ref{fig2}. The energy going into
spectral distortions is easily calculated by doing a Taylor series expansion of
the blackbody photon occupation number with position dependent temperature
up to second order and doing an ensemble average. The result for energy in
$y$,$i$ and $\mu$-type distortions ($\frac{\id \mu}{\id t}=1.4\frac{\id}{\id t}\frac{\Delta E}{E_{\gamma}}$), $\Delta E/E_{\gamma}$, is   \cite{cks2012,ksc2012b}
\begin{align}
\frac{\id}{\id t}\frac{\Delta E}{E_{\gamma}}&=\frac{-6}{3}\frac{\id}{\id t}\int \frac{k^2\id k}{2\pi^2}P_i(k)\left[\sum_{\ell=0}^{\infty}(2\ell+1)\Theta
_{\ell}^2\right]
 \approx -2\frac{\id}{\id t}\int \frac{k^2\id k}{2\pi^2}P_i(k)\left[\Theta
_{0}^2+3\Theta_1^2\right],\nonumber
\end{align}
where $\Theta_{\ell}(k)$ are the multipole moments of spherical harmonic
decomposition of temperature anisotropies, and $P_i(k)$ is the initial power
spectrum. Before recombination the above expression is easily calculated
using the tight coupling solutions\cite{hs1995}. In particular, as a check, it should be
noted that the oscillating part of the  monopole ($\propto \cos(kr_s)$,
where $r_s$ is the sound horizon) and
dipole ($\propto \sin(kr_s)$) terms in the tight coupling limit in the
radiation dominated era have same amplitude but are out of phase by $\pi/2$. The total energy in the
sound waves is conserved and  it is only the kinetic and internal energy
parts which oscillate, as expected.  There is, thus,  no need to do
the traditional 'averaging over
an oscillation' when calculating the sound wave dissipation. \footnote{Thanks to Nail Inogamov for discussion on this
    aspect.}  Finally, the shape of the $i$-type distortion depends on the spectral
  index $\nS$ of the primordial power spectrum and thus provides a  way to
  measure $\nS$  on small scales.\cite{ks2012b}

\bibliographystyle{ws-procs975x65}
\bibliography{main}

\begin{thebibliography}{10}

\bibitem{cobe}
{D. J. Fixsen et al.}, {\em \apj} {\bf 473}, p. 576 (1996).

\bibitem{cobedmr}
{E. L. Wright et al.}, {\em \apjl} {\bf 464}, p. L21 (1996).

\bibitem{wmap}
{D. Larson et al.}, {\em \apjs} {\bf 192}, p.~16 (2011).

\bibitem{spt}
{R. Keisler et al.}, {\em \apj} {\bf 743}, p.~28 (2011).

\bibitem{act}
{R. Hlozek et al.}, {\em \apj} {\bf 749}, p.~90 (2012).

\bibitem{ly2006}
{P. McDonald et al.}, {\em \apjs} {\bf 163}, 80 (2006).

\bibitem{ssm2006}
U.~{Seljak}, A.~{Slosar} and P.~{McDonald}, {\em \jcap} {\bf 10}, p.~14 (2006).

\bibitem{sz1970}
R.~A. {Sunyaev} and Y.~B. {Zeldovich}, {\em \apss} {\bf 7}, 20 (1970).

\bibitem{dd1982}
L.~{Danese} and G.~{de Zotti}, {\em \aap} {\bf 107}, 39 (1982).

\bibitem{ks2012b}
R.~{Khatri} and R.~A. {Sunyaev}, {\em \jcap} {\bf 9}, p.~16 (2012).

\bibitem{zs1969}
Y.~B. {Zeldovich} and R.~A. {Sunyaev}, {\em \apss} {\bf 4}, 301 (1969).

\bibitem{silk}
J.~{Silk}, {\em ApJ} {\bf 151}, p. 459 (1968).

\bibitem{Peebles1970}
P.~J.~E. {Peebles} and J.~T. {Yu}, {\em \apj} {\bf 162}, p. 815 (1970).

\bibitem{kaiser}
N.~{Kaiser}, {\em MNRAS} {\bf 202}, 1169 (1983).

\bibitem{cks2012}
J.~{Chluba}, R.~{Khatri} and R.~A. {Sunyaev}, {\em \mnras} {\bf 425}, 1129
  (2012).

\bibitem{pixie}
{A. Kogut et al.}, {\em \jcap} {\bf 7}, p.~25 (2011).

\bibitem{sz1970b}
R.~A. {Sunyaev} and Y.~B. {Zeldovich}, {\em Ap\&SS} {\bf 9}, 368 (1970).

\bibitem{hss94}
W.~{Hu}, D.~{Scott} and J.~{Silk}, {\em ApJl} {\bf 430}, L5 (1994).

\bibitem{daly1991}
R.~A. {Daly}, {\em \apj} {\bf 371}, 14 (1991).

\bibitem{zis1972}
Y.~B. {Zeldovich}, A.~F. {Illarionov} and R.~A. {Sunyaev}, {\em Soviet JETP}
  {\bf 35}, p. 643 (1972).

\bibitem{cs2004}
J.~{Chluba} and R.~A. {Sunyaev}, {\em \aap} {\bf 424}, 389 (2004).

\bibitem{ksc2012b}
R.~{Khatri}, R.~A. {Sunyaev} and J.~{Chluba}, {\em \aap} {\bf 543}, p. A136
  (2012).

\bibitem{hs1995}
W.~{Hu} and N.~{Sugiyama}, {\em \apj} {\bf 444}, 489 (1995).

\end{thebibliography}

\end{document}